\documentclass[aps,preprint,nofootinbib,eqsecnum]{revtex4}
\usepackage{amsmath}
\usepackage{amsfonts}
\usepackage{amssymb}
\usepackage{graphicx}
\usepackage{amscd}

 \allowdisplaybreaks

\begin{document}

\begin{flushleft}
USCHEP/0308ib3
\hfill hep-th/0308107 \\
UCB-PTH-03/20 \hfill CERN-CH/2003-188 \\
LBNL-53577
\end{flushleft}

\title{Superstar in Noncommutative Superspace\\
via Covariant Quantization of the Superparticle }
\author{Itzhak Bars}
\affiliation{Department of Physics, USC, Los Angeles, CA 90089-0484, USA \\
Theory Division, CERN, CH-1211 Geneva 23, Switzerland}
\author{Cemsinan Deliduman }
\affiliation{Feza G\"{u}rsey Institute, \c{C}engelk\"{o}y 81220, \.{I}stanbul, Turkey}
\author{Andrea Pasqua and Bruno Zumino}
\affiliation{Lawrence Berkeley National Laboratory, 1 Cyclotron Rd., Berkeley, CA 94720,
USA}

\begin{abstract}
A covariant quantization method is developed for the off-shell
superparticle in 10 dimensions. On-shell it is consistent with
lightcone quantization, while off-shell it gives a noncommutative
superspace that realizes non-linearly a hidden 11-dimensional
super Poincar\'{e} symmetry. The non-linear commutation rules are
then used to construct the supersymmetric generalization of the
covariant Moyal star product in noncommutative superspace. As one
of the possible applications, we propose this new product as the
star product in supersymmetric string field theory. Furthermore,
the formalism introduces new techniques and concepts in
noncommutative (super)geometry.
\tableofcontents
\end{abstract}

\maketitle


\section{Motivation: Star Product in Superstring Field Theory}

The purpose of this note is to propose the covariant spacetime
supersymmetric generalization of the Moyal star product as a first step in
constructing supersymmetric string field theory. The motivation for this
work is provided by the Moyal star formulation of string field theory (MSFT)
\cite{B1}-\cite{DLMZ}. On the way to constructing the superstar, we also
obtain new results on the quantization of the off-shell superparticle, and
on new group theoretical methods for constructing and evaluating star
products based on nontrivial (super)Poisson manifolds.

The first proposal of a covariant nontrivial product in superspace was given
in the context of purely fermionic supergravity \cite{BaMac}, as $\theta
^{\alpha }\cdot \theta ^{\beta }=C^{\alpha \beta }$, where $C^{\alpha \beta
} $ is the charge conjugation matrix\footnote{$C^{\alpha \beta }$ is
antisymmetric for $d=(3,4,5)\ mod(8)$, symmetric for $d=(7,8,9)\ mod(8)$ and
mixed (i.e. Lorentz singlet occurs in product of opposite chiral spinors) in
$d=(6,10)\ mod(8)$. The product $\theta ^{\alpha }\cdot \theta ^{\beta
}=C^{\alpha \beta }$ is not associative, but an associative \textit{covariant%
} fermionic Moyal product $\theta ^{\alpha }\star \theta ^{\beta }=\theta
^{\alpha }\theta ^{\beta }+\frac{1}{2}C^{\alpha \beta }$ can be constructed
in every dimension generally as $A(\theta )\star B(\theta )$, where $\star
=\exp \left( -\frac{1}{2}C^{\alpha \beta }\frac{\overleftarrow{\partial }}{%
\partial \theta ^{\alpha }}\frac{\overrightarrow{\partial }}{\partial \theta
^{\beta }}\right) $, because the star anticommutator $\{\theta ^{\alpha
},\theta ^{\beta }\}_{\star }$ is either zero (antisymmetric $C^{\alpha
\beta }$) or a constant (symmetric or mixed $C^{\alpha \beta }$).}. A later
proposal was given in \cite{VanSch} as $\left\{ \theta _{\alpha },\theta
_{\beta }\right\} =x^{\mu }\left( \gamma _{\mu }\right) _{\alpha \beta }.$
These ideas were motivated by certain aspects of supergravity or
supersymmetry and their mysterious origins were not at that time connected
to string theory. In the recent literature there are other studies of a star
product in non-covariant superspace \cite{VG}\cite{oogurivafa}\cite{seiberg}%
(see also \cite{balachandran}\cite{klemm}) whose origin is
background fields in string theory. The star product in MSFT has
also a fundamental but different physical origin, namely string
joining/splitting. The superstar product we study in this paper is
motivated by MSFT, and as required in that context, is super
Poincar\'{e} invariant, and has a different structure than the
previous proposals.

It has been shown that in the language of string field theory the Moyal
product is the simplest description of interactions of bosonic strings,
corresponding to string joining or splitting \cite{B1}. To arrive at this
description we express the general string field in the space of mixed
position-momentum representation of string modes $A\left( \bar{x}%
,x_{e},p_{e}\right) $ (instead of purely position representation), where $%
\bar{x}^{\mu }$ is the string midpoint, and $\left( x_{e}^{\mu },p_{e}^{\mu
}\right) ,$ with $e=2,4,6,\cdots $, is an equivalent description of the
string excitation modes, that are compatible with simultaneous observations
in first quantized quantum mechanics of the string\footnote{%
The probability amplitude in position space is $A\left( \bar{x}%
,x_{e},x_{o}\right) \equiv <\bar{x},x_{e},x_{o}|A>$ where $o=1,3,5,\cdots $
and $e=2,4,6,\cdots $ denote excited modes and $\bar{x}$ is the midpoint
mode. In the mixed even positions and odd momenta space (obtained by Fourier
transformation) the probability amplitude is $A\left( \bar{x}%
,x_{e},p_{o}\right) \equiv <\bar{x},x_{e},p_{o}|A>$. As in \cite{B1} we
define $p_{e}$ as a linear combination of the odd momentum modes $%
p_{e}=\sum_{o}p_{o}R_{oe}$ leading to the probability amplitude $<\bar{x}%
,x_{e},p_{e}|A>=A\left( \bar{x},x_{e},p_{e}\right) .$ It is important to
emphasize that here $p_{e}$ is \textit{not} the momentum that is quantum
canonical conjugate to $x_{e}$ as defined in the canonical treatment of
string modes. That mode is represented by $-i\partial _{x_{e}}$ as applied
on the string field $A\left( \bar{x},x_{e},p_{e}\right) $. Instead, $p_{e}$
is defined as a linear combination of the odd momentum modes as above. Since
$x_{e}$ and $p_{o}$ commute in quantum mechanics, $x_{e}$ and $p_{e}$ also
commute with each other in quantum mechanics, and therefore $\left(
x_{e},p_{e}\right) $ are quantum mechanically \textit{compatible
observables, }as they should be in defining the probability amplitude. At
first sight it appears somewhat misleading to use the symbol $p_{e},$ since
that symbol is also used for the \textit{quantum} canonical conjugate to $%
x_{e}$. However, the string joining/splitting operation introduces a new
noncommutativity beyond quantum mechanics, such that $x_{e}^{\mu
},p_{e}^{\mu }$ behave like a canonical pair under the string
joining/splitting star product as in Eq.(\ref{relstar}), although $%
x_{e},p_{e}$ commute with each other in quantum mechanics. In this sense the
usual momentum mode $-i\partial _{x_{e}}A$ is actually reproduced as a
star-commutator $-i\partial _{x_{e}}A=\left[ p_{e},A\right] _{\star },$ and
therefore, after all, $p_{e}$ does behave as if it is a canonical conjugate
to $x_{e},$ justifying the use of the symbol $p_{e}.$ \label{pe}}. Then the
joining of strings is described by combining their first-quantized
probability amplitudes (i.e. string fields) $A_{1}\left( \bar{x}%
,x_{e},p_{e}\right) $ and $A_{2}\left( \bar{x},x_{e},p_{e}\right) $ into the
probability amplitude $A_{12}\left( \bar{x},x_{e},p_{e}\right) $%
\begin{equation}
A_{12}\left( \bar{x},x_{e},p_{e}\right) =A_{1}\left( \bar{x}%
,x_{e},p_{e}\right) \star A_{2}\left( \bar{x},x_{e},p_{e}\right) ,
\end{equation}%
where the star product is local at the midpoint $\bar{x}^{\mu }$ of both
strings, and is precisely the Moyal product in the noncommutative space $%
x_{e},p_{e},$ separately for each $e=2,4,6,\cdots ,$ as in \cite{B1}%
\begin{equation}
\star \equiv \prod\limits_{e\geq 2}e^{\frac{i}{2}\left( \overleftarrow{%
\partial }_{x_{e}}\overrightarrow{\partial }_{p_{e}}-\overleftarrow{\partial
}_{p_{e}}\overrightarrow{\partial }_{x_{e}}\right) }.
\end{equation}%
This formulation of string interactions, which reproduces the operator
formalism or conformal field theory, has developed into an efficient
computational tool in string field theory \cite{B1}-\cite{DLMZ}.

The meaning of the symbol $p_{e}^{\mu }$ must be emphasized. Namely, as
described in footnote (\ref{pe}), it must be clearly understood that the
canonical-conjugate-like properties of $p_{e}$ in MSFT
\begin{equation}
x_{e}^{\mu }\star p_{e^{\prime }}^{\nu }-p_{e^{\prime }}^{\nu }\star
x_{e}^{\mu }=i\delta _{ee^{\prime }}\eta ^{\mu \nu },  \label{relstar}
\end{equation}%
derive from string joining/splitting, and not from quantum mechanics%
\footnote{%
However, this observation leads us to speculate that the mysterious origin
of quantum mechanics may be related to some deeper physical phenomenon,
analogous to string joining/splitting.}. Thus the process of string
joining/splitting creates the noncommutative space $\left( x_{e}^{\mu
},p_{e}^{\mu }\right) $ which includes timelike coordinates and is Lorentz
covariant. The ghost problems of the timelike coordinates is taken care of
by the overall gauge invariance structure of MSFT.

Since the star product above is independent for each string mode, we may
concentrate on the supersymmetrization of the Moyal product for one degree
of freedom. To do this we are inspired by the close relation between the
string joining/splitting star product for a single mode as given in Eq.(\ref%
{relstar}), and the quantum mechanics of a single relativistic particle
which has exactly the same mathematical structure. That is, to
supersymmetrize the string joining/splitting star product we will borrow
from the supersymmetry structure of the quantum mechanics of the
relativistic supersymmetric particle.

Thus we consider the phase space degrees of freedom of a single relativistic
superparticle given by $\left( x^{\mu },p^{\mu },\theta ^{\alpha },\pi
_{\alpha }\right) ,$ where $\mu $ denotes the vector and $\alpha $ denotes
the spinor in $d$ dimensions, with $\left( x^{\mu },p^{\mu }\right) $ and $%
\left( \theta ^{\alpha },\pi _{\alpha }\right) $ being canonical conjugates.
However, there is also a fermionic constraint such that $\pi _{\alpha }$ is
not an independent degree of freedom, and results in the supercharge being
proportional to $\theta $%
\begin{equation}
Q_{\alpha }\sim \left( \not{p}\theta \right) _{\alpha },\;\left( \not{p}%
\right) _{\alpha \beta }\equiv p_{\mu }\left( \gamma ^{\mu }\right) _{\alpha
\beta }.  \label{Qtheta}
\end{equation}%
Hence we may take only $\left( x^{\mu },p^{\mu },Q_{\alpha }\right) $ as the
independent degrees of freedom, and consider fields in super phase space of
the form $A\left( x,p,Q\right) .$ We can think of these functions as the
Weyl images of corresponding operators in quantum mechanics, and define a
star product among them such that the star product among the Weyl images
reproduces the products of the operators in quantum mechanics. The star
product thus defined is the generalization of the Moyal product which is
automatically invariant under relativistic supersymmetry transformations. We
then propose the same mathematical structure as the supersymmetrization of
string joining/splitting one mode at a time, generalizing the product in Eq.(%
\ref{relstar}).

In the present paper we discuss the superparticle and the corresponding star
product in its own right. It remains to be examined in the future whether
this proposal for supersymmetrizing one mode at a time really reproduces the
joining/splitting of superstrings.

Thus we will first propose a novel covariant quantization of the off-shell
superparticle in section 2, and then study the star product in the
noncommutative covariant superspace that emerges, in section 3. In this
approach to quantization of the superparticle, we will deviate from the
structure of the superparticle in one respect, namely we will not impose the
mass shell condition $p^{2}=0$ which also implies the constraint $\not{p}Q=0$%
. These constraints will be relaxed because the string modes which we wish
to consider are off shell and do not satisfy these conditions. Then we find
that the quantum theory of the off-shell 10-dimensional superparticle is
described by a \textit{non-linear realization} of the 11-dimensional
Poincar\'e superalgebra. The superspace thus defined is non-commutative, and
it becomes the basis for our proposal for the star product in supersymmetric
string field theory.

\section{Covariant Quantization of the Off-shell Superparticle}

The standard generator of supersymmetry acting on \textit{unconstrained}
super phase space is $Q_{\alpha }=\pi _{\alpha}+\left( \not{p}\theta \right)
_{\alpha }.$ The commutation rules among $Q_{\alpha }$ and other functions
of phase space follow from the canonical commutation rules $\left[ x_{\mu
},p_{\nu }\right] =i\eta _{\mu \nu }$ and $\left\{ \pi _{a},\theta ^{\beta
}\right\} =\delta _{\alpha }^{\beta }.$ In particular $Q_{\alpha }$ and $%
p_{\mu }$ satisfy the standard Poincar\'{e} superalgebra%
\begin{equation}
\left\{ Q_{\alpha },Q_{\beta }\right\} =2\left( \not{p}\right) _{\alpha
\beta },\;\left[ Q_{\alpha },p_{\mu }\right] =0,\;\left[ p_{\mu },p_{\nu }%
\right] =0.  \label{salgebra}
\end{equation}

The superparticle is defined with a constrained super phase space.
In particular, one finds that the following combination of
canonical variables vanishes $d_{\alpha }\equiv $ $\pi _{\alpha
}-\left( \not{p}\theta \right) _{\alpha }=0.$ Therefore, in the
subspace $\pi _{\alpha }=\left( \not{p}\theta \right) _{\alpha }$
the supercharge $Q_{\alpha }$ takes the form $Q_{\alpha}\sim
\left( \not{p}\theta \right) _{\alpha }.$ Due to such relations
the independent degrees of freedom need to be identified and then
the correct commutation rules need to be worked out for the
constrained subspace. Whatever these may turn out to be for some
chosen independent degrees of freedom, the supersymmetry algebra
of Eq.(\ref{salgebra}) must remain unchanged even for the
constrained system, because this algebra is a reflection of the
supersymmetry of the theory.

In non-covariant quantization, such as in the lightcone gauge, there is no
problem in identifying and quantizing the independent degrees of freedom%
\footnote{%
In the lightcone gauge one chooses $\gamma ^{+}\theta =0$ and $%
x^{+}=p^{+}\tau ,$ and the mass shell condition $p^{2}=0$ is solved by $%
p^{-}=\left( p^{i}\right) ^{2}/\left( 2p^{+}\right) .$ In 10 dimensions, the
remaining independent degrees of freedom are $\left(
x^{-},x^{i};p^{+},p^{i},\chi _{a}\right) $ where $i=1,\cdots ,8$ labels SO$%
\left( 8\right) $ vectors and $a=1,\cdots ,8$ labels SO$\left( 8\right) $
spinors. Using the following 16$\times $16 gamma matrix representation in
the lightcone basis $\gamma ^{+}=\left(
\begin{array}{cc}
0 & \sqrt{2} \\
0 & 0%
\end{array}%
\right) ,\;\gamma ^{-}=\left(
\begin{array}{cc}
0 & 0 \\
-\sqrt{2} & 0%
\end{array}%
\right) ,\;\gamma ^{i}=\left(
\begin{array}{cc}
\sigma ^{i} & 0 \\
0 & -\sigma ^{i}%
\end{array}%
\right) ,$ we can write $\not{p}=-p^{+}\gamma ^{-}-p^{-}\gamma
^{+}+p^{i}\gamma ^{i},$ and obtain the gauge fixed forms $\theta =\sqrt{%
\sqrt{2}/p^{+}}\left(
\begin{array}{c}
\chi \\
0%
\end{array}%
\right) $ and $Q=\not{p}\theta =\sqrt{\sqrt{2}/p^{+}}\left(
\begin{array}{c}
p^{i}\sigma ^{i}\chi \\
\sqrt{2}p^{+}\chi%
\end{array}%
\right) $. We can then show that the basic supersymmetry algebra in Eq.(\ref%
{salgebra}) follows from the commutation rules of the independent canonical
variables $\left( x^{-},x^{i};p^{+},p^{i},\chi _{a}\right) $ given by $\left[
x^{-},p^{+}\right] =-i,$ $\left[ x^{i},p^{j}\right] =i\delta ^{ij},$ $%
\left\{ \chi _{a},\chi _{b}\right\} =\delta _{ab}.\label{lightcone}$} of the
massless, on shell $p^{2}=0,$ superparticle \cite{brinkschwarz}. However,
the covariant quantization of the massless superparticle has been a
longstanding problem. Despite many attempts in a variety of approaches that
provide an answer consistent with lightcone quantization, there is still
room for discussion \cite{berkoParticle}\cite{vannieu} of what is an
economical approach to covariant quantization. This problem has attracted a
lot of attention because it is a first step toward the much harder problem
of covariant quantization of the superstring in the Green-Schwarz formalism
\cite{berkoString}.

In our investigation in this paper we will relax the mass shell
condition and allow any value for $p^{2}.$ This is a desirable
step anyway for massless superparticles which are off shell in the
presence of interactions. It is also desirable for the application
we have in mind in string field theory as explained in the
previous section. Since the constraint due to kappa supersymmetry
$\not{p}Q=0$ is not satisfied off-shell, the fermionic gauge
symmetry is no longer effective. Hence the off-shell superparticle
has more degrees of freedom. In this case we see that, at least
heuristically, we can solve for $\theta $ from the constraint
$\theta
_{\alpha }\sim \left( \not{p}^{-1}Q\right) _{\alpha },$ where $\left( \not{p}%
^{-1}\right) _{\alpha \beta }=\frac{1}{p^{2}}\not{p}_{\alpha \beta },$ so
that we may attempt to formulate the quantum theory covariantly in terms of
the off-shell independent degrees of freedom $\left( Q_{\alpha },p_{\mu
},x_{\mu }\right) ,$ while treating $\theta $ as a dependent quantity. We
preferred to eliminate $\theta $ and keep $Q$ as the independent dynamical
quantity since the commutation relations of $Q$ with any quantity have the
meaning of infinitesimal supersymmetry transformations and therefore its
commutators can be obtained from SUSY transformations. In particular we
already know the quantum algebra of $\left( Q,p\right) $ through the SUSY
algebra in Eq.(\ref{salgebra}). As we have already emphasized, the SUSY
algebra must be obeyed in any procedure of quantization because of
consistency with the underlying global symmetry of the theory.

What remains is to find the commutation rules of $x^{\mu }.$ In particular
we need to find $\left[ Q_{\alpha },x^{\mu }\right] $ and $\left[ x^{\mu
},x^{\nu }\right] .$ These are generally gauge dependent since $x^{\mu }$
transforms under the fermionic local symmetry as well as under the
reparametrization symmetry. To find these we will require consistency with
the covariant canonical commutation rule
\begin{equation}
\left[ x^{\mu },p^{\nu }\right] =i\eta ^{\mu \nu },  \label{xp}
\end{equation}%
and with the covariant SUSY algebra of Eq.(\ref{salgebra}).

Since $\left[ Q_{\alpha },x^{\mu }\right] $ amounts to an infinitesimal SUSY
transformation, we recall that, for unconstrained superspace it was given as
$-i\left( \gamma ^{\mu }\theta \right) _{\alpha }.$ Using this as a hint,
and noting that $\theta \sim \left( \not{p}^{-1}Q\right) ,$ we write $\left[
Q_{\alpha },x^{\mu }\right] =-ic\left( \gamma ^{\mu }\not{p}^{-1}Q\right)
_{\alpha }$ where we determine the unknown constant $c$ by consistency with
Jacobi identities. Specifically, the super Jacobi identity for $\left(
Q_{\alpha },Q_{\beta },x^{\mu }\right) =0,$ together with Eqs.(\ref{salgebra}%
,\ref{xp}) determine $c=1/2.$ Therefore we find%
\begin{equation}
\left[ Q_{\alpha },x^{\mu }\right] =-\frac{i}{2}\left( \gamma ^{\mu }\not{p}%
^{-1}Q\right) _{\alpha }.  \label{Qx}
\end{equation}%
Next we examine the Jacobi identity for $\left( x^{\mu },x^{\nu },Q_{\alpha
}\right) =0$ to find the commutator $\left[ x^{\mu },x^{\nu }\right] .$ We
can easily see that this commutator cannot vanish, and therefore we must
have a noncommutative space $x^{\mu }$. Using symmetry/antisymmetry
properties of gamma matrices in 10 dimensions, Lorentz covariance, and
dimensions of operators $(\frac{1}{2},1,-1)$ respectively for $\left(
Q,p,x\right) $, we can guess the only possible structure to be $\left[
x^{\mu },x^{\nu }\right] =\frac{b}{\left( p^{2}\right) ^{2}}Q\left\{ \gamma
^{\mu \nu },\not{p}\right\} Q$ up to the unknown constant $b$. Imposing the
Jacobi identity $\left( x^{\mu },x^{\nu },Q_{\alpha }\right) =0$ we find $%
b=-1/16.$ Therefore, we have%
\begin{equation}
\left[ x^{\mu },x^{\nu }\right] =-i\frac{S^{\mu \nu }}{p^{2}},\;\;S^{\mu \nu
}\equiv \frac{-i}{16p^{2}}Q\left\{ \gamma ^{\mu \nu },\not{p}\right\} Q.
\label{xx}
\end{equation}%
This noncommutative algebra among the $x^{\mu }$ is consistent with the
commutative subspace in the lightcone gauge, as seen from $\not{p}%
Q\rightarrow 0$ if one uses the lightcone form\footnote{%
Due to powers of $p^{2}$ in the denominator, the on-shell massless particle
condition $p^{2}=0$ leads to ambiguous expressions $0/0.$ However, the
lightcone massless particle is \textit{defined }by eliminating some of its
degrees of freedom through gauge conditions, and this procedure requires us
to interpret $0/0$ as zero when comparing to the lightcone quantization of
the massless particle.} of $Q$ and $\not{p}$ in footnote (\ref{lightcone}).

In preparation for the Jacobi identity among three $x^{\mu }$'s we evaluate%
\begin{equation}
\left[ \left[ x^{\mu },x^{\nu }\right] ,x^{\lambda }\right] =\frac{1}{p^{4}}%
\left( p^{\mu }S^{\lambda \nu }-p^{\nu }S^{\lambda \mu }+2p^{\lambda }S^{\mu
\nu }\right) .
\end{equation}%
To arrive at this form we used a number of gamma matrix identities, and the
form $Q\gamma ^{\mu }Q=p^{\mu }$ which follows from the symmetric $\left(
\gamma ^{\mu }\right) _{\alpha \beta }$ and the anticommutator in Eq.(\ref%
{salgebra}). From this, it is immediately seen that the Jacobi identity is
satisfied $\left( x^{\mu },x^{\nu },x^{\lambda }\right) =0$. All other
Jacobi identities among the quantities $Q_{\alpha },p^{\mu },x^{\mu }$ are
trivially satisfied.

Hence we have shown that the off-shell covariant quantization of the
superparticle is uniquely determined by the basic commutations rules in Eqs.(%
\ref{salgebra}-\ref{xx}). In our covariant quantization approach we were
guided only by the consistency with the global symmetry. The consistent
non-linear algebra defined by these equations is a nonlinear extension of
the well known SUSY algebra of Eq.(\ref{salgebra}).

The results above can also be derived from the Dirac brackets in
covariant quantization as follows. The constraints $d_{\alpha
}\equiv \pi _{\alpha}-\left( \not{p}\theta \right) _{\alpha }$
obey the algebra $\left\{ d_{\alpha },d_{\beta }\right\}
=-2\not{p}_{\alpha \beta }.$ When $p^{\mu }$
is off shell these are second class constraints. Using the constraints $%
d_{\alpha },$ one can compute the classical Dirac brackets among $\left(
x^{\mu },p_{\mu },\theta ^{\alpha },\pi _{\alpha }\right) $ as follows%
\begin{eqnarray}
\left\{ x^{\mu },p_{\nu }\right\} _{D} &=&\delta _{\nu }^{\mu },\;\;\left\{
x^{\mu },\theta ^{\alpha }\right\} _{D}=-\frac{1}{2}\left( \not{p}%
^{-1}\gamma ^{\mu }\theta \right) ^{\alpha },\;\;\left\{ x^{\mu
},\pi _{\alpha}\right\} _{D}=\frac{1}{2}\left( \gamma ^{\mu
}\theta \right) _{\alpha
},\;\; \\
\left\{ \theta ^{\alpha },\theta ^{\beta }\right\} _{D} &=&\frac{1}{2}\left(
\not{p}^{-1}\right) ^{\alpha \beta },\;\;\left\{ \theta ^{\alpha },\pi
_{\beta }\right\} _{D}=\frac{1}{2}\delta _{\beta }^{\alpha },\;\;\left\{ \pi
_{\alpha },\pi _{\beta }\right\} _{D}=\frac{1}{2}\not{p}_{\alpha \beta },\;\;
\\
\left\{ x^{\mu },x^{\nu }\right\} _{D} &=&\frac{i}{4}\theta \left\{ \gamma
^{\mu \nu },\not{p}^{-1}\right\} \theta .
\end{eqnarray}%
Solving the constraints one can write $\pi _{\alpha }=\left( \not{p}\theta
\right) _{\alpha },$ $Q_{\alpha }=2\pi _{\alpha}$
and $\theta _{\alpha }=\frac{1}{%
2}\left( \not{p}^{-1}Q\right) _{\alpha }.$ Eliminating $\theta $
and $\pi $ in favor of $Q$ through these equations, and inserting
the factor of $i$ in passing to quantum mechanics, we arrive at
the same relations derived above through the Jacobi identities.

Next, we examine further the properties of $S^{\mu \nu }.$ We see that it
commutes with the momentum $p^{\mu },$ it is transverse to it $S^{\mu \nu
}p_{\nu }=0,$ and from Eq.(\ref{salgebra}) it follows that it satisfies the
algebra of Lorentz transformations in the space transverse to $p^{\mu }$%
\begin{equation}
\left[ S^{\mu \nu },S^{\lambda \sigma }\right] =i\left( S^{\mu \lambda
}\left( \eta ^{\nu \sigma }-\frac{p^{\nu }p^{\sigma }}{p^{2}}\right) -\left(
\mu \leftrightarrow \nu \right) \right) -i\left( \left( \lambda
\leftrightarrow \sigma \right) \right) .
\end{equation}%
Therefore, $S^{\mu \nu }$ is interpreted as the spin operator. Indeed, its
commutator with $Q_{\alpha },$ as follows from Eq.(\ref{salgebra}), gives
the correct Lorentz transformation of the spinor in the subspace transverse
to $p^{\mu }$%
\begin{equation}
\left[ S^{\mu \nu },Q_{\alpha }\right] =\frac{i}{2}\left( \left( \gamma
^{\mu \nu }-\gamma ^{\mu \sigma }\frac{p^{\nu }p_{\sigma }}{p^{2}}+\gamma
^{\nu \sigma }\frac{p^{\mu }p_{\sigma }}{p^{2}}\right) Q\right) _{\alpha }.
\end{equation}

These observations lead us to introduce the following dimensionless
hermitian vector $J^{\mu }$%
\begin{equation}
J^{\mu }\equiv \left( -p^{2}\right) ^{\frac{1}{4}}x^{\mu }\left(
-p^{2}\right) ^{\frac{1}{4}}=\left( -p^{2}\right) ^{\frac{1}{2}}x^{\mu }-%
\frac{i}{2}\left( -p^{2}\right) ^{-\frac{1}{2}}p^{\mu }=x^{\mu }\left(
-p^{2}\right) ^{\frac{1}{2}}+\frac{i}{2}\left( -p^{2}\right) ^{-\frac{1}{2}%
}p^{\mu }.
\end{equation}%
We find that its commutators give the total Lorentz generator $J^{\mu \nu }$%
\begin{equation}
\left[ J^{\mu },J^{\nu }\right] =iJ^{\mu \nu },\;\;\;J^{\mu \nu }=\left(
x^{\mu }p^{\nu }-x^{\nu }p^{\mu }\right) +S^{\mu \nu }.
\end{equation}%
It is now straightforward to notice that the operators $J^{\mu \nu },p^{\mu
},Q_{\alpha }$ satisfy the super Poincar\'{e} algebra. Namely $J_{\mu \nu }$
rotates correctly $p^{\mu }$ as well as $Q_{\alpha },$ and it satisfies the
Lorentz algebra not only in the space transverse to $p^{\mu },$ but in the
full space%
\begin{eqnarray}
\left[ J^{\mu \nu },p^{\lambda }\right] &=&i\eta ^{\lambda \lbrack \mu
}p^{\nu ]},\;\left[ J^{\mu \nu },Q_{\alpha }\right] =\frac{i}{2}\left(
\gamma ^{\mu \nu }Q\right) _{\alpha }  \label{susy10-1} \\
\left[ J^{\mu \nu },J^{\lambda \sigma }\right] &=&i\left( J^{\mu \lambda
}\eta ^{\nu \sigma }-\left( \mu \leftrightarrow \nu \right) \right) -i\left(
\lambda \leftrightarrow \sigma \right) .  \label{susy10-2}
\end{eqnarray}%
Hence our covariant quantization of the off shell superparticle is
consistent with the global Poincar\'{e} symmetry of the theory.

Furthermore, we can extend the symmetry algebra into a hidden symmetry in
11-dimensions, by including the operators $J^{\mu },$ $\left( -p^{2}\right)
^{\frac{1}{2}},$ and $\tilde{Q}_{\dot{\alpha}}\equiv \left( -p^{2}\right) ^{-%
\frac{1}{2}}\left( \not{p}Q\right) _{\dot{\alpha}}.$ We compute the
commutators of $J^{\mu }$ with the other quantities and find%
\begin{eqnarray}
\left[ J^{\mu },\left( -p^{2}\right) ^{\frac{1}{2}}\right] &=&-ip^{\mu },\;%
\left[ J^{\mu },p^{\nu }\right] =i\eta ^{\mu \nu }\left( -p^{2}\right) ^{%
\frac{1}{2}},\;\left[ J^{\mu },J^{\nu \lambda }\right] =-i\eta ^{\mu \lbrack
\nu }J^{\lambda ]}, \\
\left[ J^{\mu },Q_{\alpha }\right] &=&-\frac{i}{2}\left( \gamma ^{\mu }%
\tilde{Q}\right) _{\alpha },\;\left[ J^{\mu },\tilde{Q}_{\dot{\alpha}}\right]
=\frac{i}{2}\left( \gamma ^{\mu }Q\right) _{\dot{\alpha}}.
\end{eqnarray}%
Together with Eqs.(\ref{susy10-1},\ref{susy10-2}), we notice the structure
of the 11-dimensional SUSY algebra, such that $Q_{A}=( Q_{\alpha },%
\tilde{Q}_{\dot{\alpha}}) $ together form a 32 component spinor, $%
P^{M}=(p^{\mu },\left( -p^{2}\right) ^{\frac{1}{2}})$ together
form an 11-dimensional massless momentum, and $J^{MN}=\left(
J_{\mu \nu },J_{\mu }\right) $ together form the 11-dimensional
Lorentz algebra. Furthermore, if we define 11-dimensional gamma
matrices $\Gamma ^{M},$ we can check explicitly that $P^{M}\left(
\Gamma _{M}\right) _{A}^{~B}Q_{B}=0$ as well as $P^{M}P_{M}=0,$
therefore our structure corresponds to the quantum massless
superparticle in 11-dimensions. The quantum states of this
off-shell system are precisely the supergravity multiplet in
11-dimensions, but dimensionally reduced to the 10-dimensional
type-IIA supergravity multiplet.

Thus, our covariant quantization of the off-shell 10 dimensional
superparticle is described by a non-linear realization of the 11-dimensional
super Poincar\'{e} algebra, acting on the quantum states that correspond to
the 11-dimensional supergravity multiplet. The fundamental commutators of
this structure are given by Eqs.(\ref{salgebra}-\ref{xx}), including the
noncommutative position space of Eq.(\ref{xx}). In particular the
commutation rules with the operator $Q_{\alpha }$ correspond to the
supersymmetry transformations of the fundamental super coordinates $\left(
x^{\mu },p^{\mu },Q_{\alpha }\right) ,$ and this gives the supersymmetry
transformation rules of fields defined as functions of this superspace $%
A\left( x,p,Q\right) .$

In our discussion we treated the 10-dimensional superparticle, but it is
straightforward to apply the same approach in any dimension $d$, leading to
a non-linear realization of the Poincar\'e superalgebra in $d+1$ dimensions.

The nonlinear superalgebra among the noncommutative covariant
superspace coordinates $\left( x^{\mu },p^{\mu },Q_{\alpha
}\right) $ is the basis for constructing the supersymmetric star
product. In principle one can use the Kontsevich method
\cite{kontsevich}\cite{chepelev} to construct the
 \textit{associative} star product. However, we can use simpler methods
that can be applied to our problem as developed in the next
section. Thus, we will first discuss a generic linear case and
later apply the method to the nonlinear superalgebra in our case.

\section{Star Product in Noncommutative Superspace}

\subsection{At most linear Poisson structure}

Consider some noncommutative space $x_{a}$ with a noncommutativity function $%
\theta _{ab}\left( x\right) ,$ such that $\theta _{ab}\left( x\right) $ is
at most linear in $x$. The classical Poisson structure then has the form%
\begin{equation}
\left\{ x_{a},x_{b}\right\} =-i\theta _{ab}\left( x\right) =-i\left( \sigma
_{ab}+if_{ab}^{c}x_{c}\right) .
\end{equation}%
where $\sigma _{ab},f_{ab}^{c}$ are independent of $x.$ Such a $\theta
_{ab}\left( x\right) $ includes the case of the Heisenberg algebra (when $%
f_{ab}^{c}=0$) as well as the case of a pure Lie algebra (when $\sigma
_{ab}=0$). In the next section we will study the case of superspace with the
$x_{a}$ replaced by the phase space of a superparticle $x_{a}\rightarrow
\left( x_{\mu },p_{\mu },Q_{\alpha }\right) $ where $Q_{\alpha }$ is the
generator of supersymmetry. But in the present section we discuss the star
product for the general bosonic system that has the general noncommutativity
property given above.

The usual quantization of this system is done in quantum mechanics by
promoting the $x_{a}$ to operators $\hat{x}_{a}.$ General operators $\hat{A}%
_{1}\left( \hat{x}\right) $, $\hat{A}_{2}\left( \hat{x}\right) $ are
multiplied with each other in the usual way by writing them next to each
other $\hat{A}_{12}\left( \hat{x}\right) =\hat{A}_{1}\left( \hat{x}\right)
\hat{A}_{2}\left( \hat{x}\right) ,$ and then the resulting operator $\hat{A}%
_{12}\left( \hat{x}\right) $ is computed by keeping track of the orders of
the operators consistently with their quantum commutation relations%
\begin{equation}
\left[ \hat{x}_{a},\hat{x}_{b}\right] =\theta _{ab}\left( \hat{x}\right)
=\sigma _{ab}+if_{ab}^{c}\hat{x}_{c}.
\end{equation}

The deformation quantization of this system introduces a star product among
classical functions $A_{1}\left( x\right) ,A_{2}\left( x\right) $ to
construct a resulting classical function $A_{12}\left( x\right) $%
\begin{equation}
A_{12}\left( x\right) =A_{1}\left( x\right) \star A_{2}\left( x\right) .
\end{equation}%
The classical $A_{1}\left( x\right) ,A_{2}\left( x\right) ,A_{12}\left(
x\right) $ are \textquotedblleft Weyl images" of the corresponding operators
$\hat{A}_{1}\left( \hat{x}\right) $, $\hat{A}_{2}\left( \hat{x}\right) $, $%
\hat{A}_{12}\left( \hat{x}\right) $ and are designed to reproduce the same
results as quantum mechanics, although in the classical function $A\left(
x\right) $ the orders of $x_{a}$ do not matter. In particular, when applied
to the case of $A_{1}\left( x\right) =x_{a},~A_{2}\left( x\right) =x_{b},$
the star product should give the same result as quantum mechanics
\begin{equation}
\left[ x_{a},x_{b}\right] _{\star }\equiv x_{a}\star x_{b}-x_{b}\star
x_{a}=\theta _{ab}\left( x\right) =\sigma _{ab}+if_{ab}^{c}x_{c}.
\label{star1}
\end{equation}%
An additional property of the star product, consistent with quantum
mechanics, is that it should be associative%
\begin{equation}
A_{123}\left( x\right) =\left( A_{1}\left( x\right) \star A_{2}\left(
x\right) \right) \star A_{3}\left( x\right) =A_{1}\left( x\right) \star
\left( A_{2}\left( x\right) \star A_{3}\left( x\right) \right) .
\end{equation}

A star product with such properties can be constructed by using Kontsevich's
general diagrammatic prescription \cite{kontsevich}, which defines it as an
infinite series for an arbitrary $\theta _{ab}\left( x\right) $. However,
because of the maximum linear nature of $\theta _{ab}\left( x\right) $ there
is a much simpler closed expression which we construct in the following
section. Of course, we expect that in the limit $f_{ab}^{c}=0,$ our
expression reduces to the simple Moyal product given by $\theta
_{ab}\rightarrow \sigma _{ab}.$

\subsection{The generic star for maximum linear $\protect\theta _{ab}\left(
x\right) $}

We define the Fourier transforms of the classical functions%
\begin{equation}
A_{i}\left( x\right) =\int \left( dp\right) \tilde{A}_{i}\left( p\right)
e^{ip^{a}x_{a}}.  \label{fourier}
\end{equation}%
Constructing the star product for the Fourier basis $e^{ip_{1}^{a}x_{a}}%
\star e^{ip_{2}^{b}x_{b}},$ is equivalent to constructing it for any other
basis of functions that can be related by the Fourier transform. We recall
the Baker-Hausdorff-Campbell (BHC) theorem for quantum operators%
\begin{equation}
e^{\hat{A}}e^{\hat{B}}=e^{\hat{C}(\hat{A},\hat{B})},
\end{equation}%
where the operator $\hat{C}(\hat{A},\hat{B})$ is determined by an infinite
series given by multiple commutators%
\begin{equation}
\hat{C}(\hat{A},\hat{B})=\hat{A}+\hat{B}+\frac{1}{2}[\hat{A},\hat{B}]+\frac{1%
}{12}[[\hat{A},\hat{B}],\hat{B}]+\frac{1}{12}[\hat{A},[\hat{A},\hat{B}%
]]+\cdots
\end{equation}%
We note that this theorem relies simply on an associative product.
Therefore, the same theorem also applies to star exponentials of classical
functions $A\left( x\right) ,B\left( x\right) $ as long as we have an
associative star product
\begin{equation}
\left( e^{A\left( x\right) }\right) _{\star }\star \left( e^{B\left(
x\right) }\right) _{\star }=\left( e^{\left( A\left( x\right) +B\left(
x\right) +\frac{1}{2}\left[ A\left( x\right) ,B\left( x\right) \right]
_{\star }+\frac{1}{12}\left[ \left[ A,B\right] _{\star },B\right] _{\star }+%
\frac{1}{12}\left[ A,\left[ A,B\right] _{\star }\right] _{\star }+\cdots
\right) }\right) _{\star }.  \label{bh}
\end{equation}%
In particular, let us construct a star product such that the star
exponential of a linear function of the $x^{a}$ is equal to the classical
exponential, i.e. $\left( e^{ip\cdot x}\right) _{\star }=e^{ip\cdot x}$ for
any set of constant parameters $p^{a}.$ This relation is true for the simple
Moyal product, and we will verify that it is also true in our case, after we
give the construction of our star product. Then the BHC theorem can be
applied to the classical Fourier basis
\begin{equation}
e^{ip_{1}^{a}x_{a}}\star e^{ip_{2}^{b}x_{b}}=\exp \left(
ip_{1}^{a}x_{a}+ip_{2}^{a}x_{a}+\frac{1}{2}\left[
ip_{1}^{a}x_{a},ip_{2}^{b}x_{b}\right] _{\star }+\cdots \right)
\label{series}
\end{equation}%
The crucial observation here is that all the higher terms in $C\left(
A,B\right) \left( x\right) $ involve only star commutators, which can be
evaluated for the Fourier basis by using repeatedly Eq.(\ref{star1}). Since $%
\theta _{ab}\left( x\right) $ is at most linear in $x,$ the result is also
necessarily at most linear in $x,$ and therefore only the linear Poisson
structure is sufficient to completely evaluate the product of exponentials.
The result is identified with the classical exponential since it is designed
to be the same as the star exponential for any $p.$ Therefore, we obtain the
form%
\begin{equation}
e^{ip_{1}^{a}x_{a}}\star e^{ip_{2}^{a}x_{a}}=Z\left( p_{1},p_{2}\right) \exp
\left( ix_{a}p_{12}^{a}\right) ,
\end{equation}%
where $Z,p_{12}^{a}$ are functions of $p_{1}^{a},p_{2}^{a}$ which we
determine below. Once these functions are determined, the star product for
generic classical functions $A_{1}\left( x\right) ,A_{2}\left( x\right) $
can be given exactly in the form%
\begin{equation}
A_{12}\left( x\right) =A_{1}\left( x\right) \star A_{2}\left( x\right) =\int
\int \left( dp_{1}dp_{2}\right) \tilde{A}_{1}\left( p_{1}\right) \tilde{A}%
_{2}\left( p_{2}\right) Z\left( p_{1},p_{2}\right) \exp \left(
ix_{a}p^{a}\left( p_{1},p_{2}\right) \right) .  \label{groupstar}
\end{equation}%
This can be rewritten in terms of differentials as follows%
\begin{equation}
A_{1}\left( x\right) \star A_{2}\left( x\right) =\left[ A_{1}\left(
x_{1}\right) ~Z\left( -i\overleftarrow{\partial }_{x_{1}},-i\overrightarrow{%
\partial }_{x_{2}}\right) \exp \left( ix_{a}\Delta p^{a}\left( -i%
\overleftarrow{\partial }_{x_{1}},-i\overrightarrow{\partial }%
_{x_{2}}\right) \right) \left( A_{2}\left( x_{2}\right) \right) \right]
_{x_{1}=x_{2}=x}.  \label{stardiff}
\end{equation}%
where $\Delta p^{a}\left( p_{1},p_{2}\right) =p_{12}^{a}\left(
p_{1},p_{2}\right) -p_{1}^{a}-p_{2}^{a}.$ Note that $\Delta p^{a}$ is
multiplied by $x^{a}$ which is kept distinct from $x_{1}^{a},x_{2}^{a}$ when
the derivatives are applied. The derivative form is useful in general, but
it is particularly essential for evaluating the star products of polynomials
in $x^{a}$.

In the simplest case of $f_{ab}^{c}=0$ the infinite series in Eq.(\ref%
{series}) terminates with the first commutator $\frac{1}{2}\left[
ip_{1}^{a}x_{a},ip_{2}^{a}x_{a}\right] _{\star }=-\frac{1}{2}p_{1}^{a}\sigma
_{ab}p_{2}^{b}$ since it is independent of $x$. Similarly, for certain cases
of interest (for an example see \cite{bianca}), the series terminates after
a few terms once we reach terms that are independent of $x$. Even if the
series does not terminate, the functions $Z\left( p_{1},p_{2}\right)
,p_{12}^{a}\left( p_{1},p_{2}\right) $ can be computed by using the
properties of the Lie algebra associated with the structure constants $%
f_{ab}^{c}.$

In particular, when we specialize this expression to $f_{ab}^{c}=0,$ we have
$Z\left( p_{1},p_{2}\right) =\exp \left( -\frac{1}{2}p_{1}^{a}\sigma
_{ab}p_{2}^{b}\right) $ and $p_{12}^{a}\left( p_{1},p_{2}\right)
=p_{1}^{a}+p_{2}^{a},$ which gives
\begin{equation}
A_{12}\left( x\right) =A_{1}\left( x\right) \star A_{2}\left( x\right)
\underset{f_{ab}^{c}=0}{=}\int \int \left( dp_{1}dp_{2}\right) \tilde{A}%
_{1}\left( p_{1}\right) \tilde{A}_{2}\left( p_{2}\right) e^{-\frac{1}{2}%
p_{1}^{a}\sigma _{ab}p_{2}^{b}}e^{ix_{a}\left( p_{1}^{a}+p_{2}^{a}\right) }.
\end{equation}%
The right hand side may be written also in the derivative form $A_{1}\left(
x\right) \star A_{2}\left( x\right) =A_{1}\left( x\right) \exp \left( \frac{1%
}{2}\overleftarrow{\partial }^{a}\sigma _{ab}\overrightarrow{\partial }%
^{b}\right) A_{2}\left( x\right) ,$ which shows that the result is the usual
Moyal star product when $f_{ab}^{c}=0.$

For the general case, we compute the first few terms of the series for any $%
\sigma _{ab},f_{ab}^{c}$%
\begin{eqnarray}
\frac{1}{2}\left[ ip_{1}^{a}x_{a},ip_{2}^{b}x_{b}\right] _{\star } &=&-\frac{%
1}{2}p_{1}^{a}\sigma _{ab}p_{2}^{b}-\frac{i}{2}\left(
p_{1}^{a}p_{2}^{b}f_{ab}^{c}\right) x_{c} \\
\frac{1}{12}\left[ \left[ ip_{1}^{a}x_{a},ip_{2}^{b}x_{b}\right]
,ip_{2}^{c}x_{c}\right] &=&\frac{1}{24}\left(
p_{1}^{a}f_{ab}^{d}p_{2}^{b}\right) \sigma _{dc}p_{2}^{c}+\frac{i}{24}\left(
p_{1}^{a}p_{2}^{b}p_{2}^{c}f_{ab}^{d}f_{dc}^{e}\right) x_{e}
\end{eqnarray}%
So we obtain%
\begin{eqnarray}
Z\left( p_{1},p_{2}\right) &=&\exp \left( -\frac{1}{2}p_{1}^{a}\sigma
_{ab}p_{2}^{b}+\frac{1}{24}\left( p_{1}^{a}f_{ab}^{d}p_{2}^{b}\right) \sigma
_{dc}p_{2}^{c}+\frac{1}{24}\left( p_{2}^{a}f_{ab}^{d}p_{1}^{b}\right) \sigma
_{dc}p_{1}^{c}+\cdots \right)  \label{z} \\
p_{12}^{e}\left( p_{1},p_{2}\right) &=&p_{1}^{e}+p_{2}^{e}-\frac{1}{2}%
p_{1}^{a}f_{ab}^{e}p_{2}^{b}+\frac{1}{24}\left(
p_{1}^{a}f_{ab}^{d}p_{2}^{b}\right) f_{dc}^{e}p_{2}^{c}+\frac{1}{24}\left(
p_{2}^{a}f_{ab}^{d}p_{1}^{b}\right) f_{dc}^{e}p_{1}^{c}+\cdots  \label{p}
\end{eqnarray}%
We identify the structure of the series as follows. First, $p_{12}^{e}\left(
p_{1},p_{2}\right) $ is independent of $\sigma _{ab}$; it is fully
determined by the group multiplication property, with $p_{1}^{a},p_{2}^{b}$
being the infinitesimal group parameters associated with the Lie algebra
characterized by $f_{ab}^{c}.$ Therefore, the full series for $%
p_{12}^{e}\left( p_{1},p_{2}\right) $ can be computed from any convenient
representation of the group (see below). Second, the expression for $\ln
Z\left( p_{1},p_{2}\right) $ is completely parallel to the expression for $%
p_{12}^{e}\left( p_{1},p_{2}\right) -p_{1}^{e}-p_{2}^{e}$, except for
replacing $\sigma _{ab}$ in place of $f_{ab}^{e}$ in the last factor of each
term. In fact, $\sigma _{ab}$ may be regarded as an additional structure
constant in the centrally extended Lie algebra characterized by $f_{ab}^{e}$%
, which explains why the two series for $\ln Z\left( p_{1},p_{2}\right) $
and $p_{12}^{e}\left( p_{1},p_{2}\right) $ have a similar structure. Thus,
if $p^{e}\left( p_{1},p_{2}\right) $ is computed exactly from some
convenient group representation, $\ln Z\left( p_{1},p_{2}\right) $ can also
be computed by using its relationship to $p_{12}^{e}\left(
p_{1},p_{2}\right) $ or by using a convenient representation of the
centrally extended Lie algebra.

As an example, consider the case of $\sigma _{ab}=0$ and take $f_{ab}^{c}$
to be the structure constants for SU$\left( 2\right) .$ To compute $%
p_{12}^{a}\left( p_{1},p_{2}\right) $ exactly we use the $2\times 2$ matrix
representation $e^{ix_{a}p^{a}}\rightarrow e^{i\frac{1}{2}\sigma _{a}p^{a}}$
where $\sigma _{a}$ are the Pauli matrices. Then the matrix representation $%
e^{i\frac{1}{2}\sigma _{a}p^{a}}=\cos \frac{\left\vert p\right\vert }{2}+i%
\frac{p\cdot \sigma }{\left\vert p\right\vert }\sin \frac{\left\vert
p\right\vert }{2}$ can be used to multiply the matrices and compute an exact
expression for $p_{12}^{a}\left( p_{1},p_{2}\right) $ written as a 3-vector $%
\vec{p}$ as follows%
\begin{eqnarray}
\cos \frac{\left\vert p\right\vert }{2} &=&\cos \frac{\left\vert
p_{1}\right\vert }{2}\cos \frac{\left\vert p_{2}\right\vert }{2}-\frac{\vec{p%
}_{1}\cdot \vec{p}_{2}}{\left\vert p_{1}\right\vert \left\vert
p_{2}\right\vert }\sin \frac{\left\vert p_{1}\right\vert }{2}\sin \frac{%
\left\vert p_{2}\right\vert }{2}  \label{su21} \\
\frac{\vec{p}}{\left\vert p\right\vert }\sin \frac{\left\vert p\right\vert }{%
2} &=&\left(
\begin{array}{c}
\frac{\vec{p}_{1}}{\left\vert p_{1}\right\vert }\sin \frac{\left\vert
p_{1}\right\vert }{2}\cos \frac{\left\vert p_{2}\right\vert }{2}+\frac{\vec{p%
}_{2}}{\left\vert p_{2}\right\vert }\sin \frac{\left\vert p_{2}\right\vert }{%
2}\cos \frac{\left\vert p_{1}\right\vert }{2} \\
-\frac{\vec{p}_{1}\times \vec{p}_{2}}{\left\vert p_{1}\right\vert \left\vert
p_{2}\right\vert }\sin \frac{\left\vert p_{1}\right\vert }{2}\sin \frac{%
\left\vert p_{2}\right\vert }{2}%
\end{array}%
\right) .  \label{su22}
\end{eqnarray}%
The first equation gives the length of the vector $\left\vert p\right\vert ,$
and after inserting it in the second equation we get the full $\vec{p}$. The
expansion of this exact expression in powers of $\vec{p}_{1},\vec{p}_{2}$
reproduces the infinite series computed through the BHC theorem. Replacing
this result for $p_{12}^{a}\left( p_{1},p_{2}\right) $ in Eq.(\ref{groupstar}%
) and taking $Z\left( p_{1},p_{2}\right) =1$ gives the star product for the
case of $\sigma _{ab}=0$ and $f_{ab}^{c}$ the SU$\left( 2\right) $ structure
constants. It is a nontrivial exercise to obtain this result in the
diagramatic approach of Kontsevich.

Finally, let us verify that the star exponential is the classical
exponential $\left( e^{ip\cdot x}\right) _{\star }=e^{ip\cdot x}$
which was assumed in our approach. This would follow by showing
that the star powers are the same as the classical powers $\left(
p\cdot x\right) _{\star }^{n}=\left( p\cdot x\right) ^{n}$ for any
set of parameters $p^{a}.$ This has to be true since the dot
product $p\cdot x$ amounts to picking up a single component of $x$
in the direction of $p^{a},$ and for a single component the star
product is trivial with itself since the commutator vanishes
$\left[ p\cdot x,p\cdot x\right] _{\star }=0$. In any case one can
also verify that it is true explicitly by applying the star product of Eq.(%
\ref{stardiff}) to $\left( p\cdot x\right) ^{n_{1}}\star \left( p\cdot
x\right) ^{n_{2}}$ to show that it gives $\left( p\cdot x\right)
^{n_{1}+n_{2}}.$ This can be proven by iteration by starting with $%
n_{1}=n_{2}=1;$ and this case is easily computed by using the expansion of
the general formulas in Eqs.(\ref{z},\ref{p}).

The lesson learned in this section is that we can use the BHC theorem and
group theory to determine the star product. The case treated in this section
involved a linear Poisson structure. In the next section we will treat a
non-linear (super)Poisson structure, but the essential tool will be again
the group theoretical aspect we emphasized in this section. We will use this
concept to determine the star product in relativistic noncommutative
superspace $\left( x^{\mu },p^{\mu },Q_{\alpha }\right) .$ From the context
we will see that our approach is a more general technique than the
particular example, and therefore it can be applied more generally to other
bosonic or supersymmetric cases.

\subsection{Covariant superstar}

We will find it convenient to work with a basis of functions of $\left(
J,p,Q\right) $ instead of functions of the noncommutative phase space $%
\left( x,p,Q\right) ,$ and define the star product in the space of $\left(
J,p,Q\right) .$ This is completely general since these are related by a
change of variables\footnote{%
To be more careful, the change of variables needs to be consistent
with the corresponding quantum operators, and therefore it should
involve star multiplication of various factors. However, note
that, as can be expected, for a single power of $x^{\mu }$ or
$J^{\mu }$ the star product form is equal to its classical form
$J^{\mu }=\left( -p^{2}\right) ^{1/4}\star x^{\mu }\star \left(
-p^{2}\right) ^{1/4}=$ $x^{\mu }\left( -p^{2}\right) ^{1/2},$ or
$x^{\mu }=\left( -p^{2}\right) ^{-1/4}\diamond J^{\mu }\diamond
\left( -p^{2}\right) ^{-1/4}=$ $J^{\mu }\left( -p^{2}\right)
^{-1/2},$
without any corrections of the deformation parameter (which is set equal to $%
1$ in our formalism).} $x^{\mu }=J^{\mu }\left( -p^{2}\right) ^{-1/2}.$ The
virtue of the basis $\left( J,p,Q\right) $ is that these variables have
simple commutation rules as quantum operators (i.e. they are part of the Lie
algebra of $d+1$ dimensional super Poincare group), and this makes them
convenient for the formalism of defining a star product on classical
functions of this space. If one desires, the star product in the space of $%
\left( x,p,Q\right) $ can be extracted from the one we define. To emphasize
that the star product is defined in the $\left( J,p,Q\right) $ space we will
denote it with the symbol $\diamond ,$ while reserving the symbol $\star $
for the $\left( x,p,Q\right) $ space.

We begin with the Fourier transform as in Eq.(\ref{fourier})
\begin{equation}
A\left( J,p,Q\right) =\int dkdqd\psi ~\tilde{A}\left( k,q,\psi \right)
~e^{ik\cdot J+iq\cdot p+i\psi \cdot Q}~.  \label{fouri}
\end{equation}%
Then we need to evaluate the star product of the Fourier basis
$e^{ik_{1}\cdot J+iq_{1}\cdot p+i\psi _{1}\cdot Q}\diamond
e^{ik_{2}\cdot J+iq_{2}\cdot p+i\psi _{2}\cdot Q}$. Note that,
these equations involve conveniently a classical exponential.
Fortunately, the classical exponential of an arbitrary
\textit{linear} combination of $\left( J,p,Q\right) $ is equal to
the star exponential as long as the star product $\diamond $ is
given in the $\left( J,p,Q\right) $ basis, $ \left( e^{ik\cdot
J+iq\cdot p+i\psi \cdot Q}\right) _{\diamond }=e^{ik\cdot
J+iq\cdot p+i\psi \cdot Q}$, for any set of constants $\left(
k_{\mu },q_{\mu },\psi _{\alpha }\right) .$
This is by virtue of the fact that, as in the previous section, under the $%
\diamond $-product $\left( J,p,Q\right) $ act as generators of the Lie
algebra of the $d+1$ dimensional super Poincar\'{e} group.

We may begin to apply the BHC theorem, as in Eq.(\ref{bh}), to evaluate this
product. It becomes quickly evident that the series does not terminate since
the non-Abelian $J^{\mu \nu }$ is produced in the commutator $\left[ J^{\mu
},J^{\nu }\right] _{\diamond }.$ However, the product can be determined from
group theory since the exponential of any linear combination of generators
defines an element of the super Poincar\'{e} group in $\left( d+1\right) $%
-dimensions. Therefore, the result of the BHC series must accumulate to
become the series one obtains in group multiplication, and therefore it must
take the form of a general group element on the right hand side%
\begin{eqnarray}
&&e^{ik_{1}\cdot J+iq_{1}\cdot p+i\psi _{1}\cdot Q}\diamond e^{ik_{2}\cdot
J+iq_{2}\cdot p+i\psi _{2}\cdot Q}  \notag \\
&=&\left( e^{i\left( k_{12}\cdot J+\omega _{12}^{\mu \nu }J_{\mu \nu
}\right) +i\left( q_{12}\cdot p+z_{12}\sqrt{-p^{2}}\right) +i\left( \psi
_{12}\cdot Q+\xi _{12}\cdot \tilde{Q}\right) }\right) _{\diamond }.
\label{diamond}
\end{eqnarray}%
The important point here is that the coefficients $k_{12}^{\mu }$, $\omega
_{12}^{\mu \nu }$, $q_{12}^{\mu }$, $z_{12}$, $\left( \psi _{12}\right)
_{\alpha }$, $\left( \xi _{12}\right) _{\dot{\alpha}},$ that appear in the
exponent must be \textit{constant coefficients } (independent of $x,p,Q$).
They are functions of $\left( k_{1},k_{2};q_{1},q_{2};\psi _{1},\psi
_{2}\right) $ which can be determined from \textit{any convenient
representation of the super Poincar\'{e} group in }$d+1$ \textit{dimensions}
as in the example of Eqs.(\ref{su21},\ref{su22}). In the result we may then
replace the non-linear \textit{classical} expressions for $J_{\mu \nu },%
\tilde{Q},$
\begin{equation}
J^{\mu \nu }=\frac{J^{[\mu }p^{\nu ]}}{\sqrt{-p^{2}}}+\frac{iQ\left\{ \gamma
^{\mu \nu },\not{p}\right\} Q}{16\left( -p^{2}\right) },\;\;\tilde{Q}_{\dot{%
\alpha}}\equiv \left( -p^{2}\right) ^{-\frac{1}{2}}\left( \not{p}Q\right) _{%
\dot{\alpha}}
\end{equation}%
and $\sqrt{-p^{2}},$ thus obtaining the desired star product for the basis $%
e^{ik_{1}\cdot J+iq_{1}\cdot p+i\psi _{1}\cdot Q}.$

Note that, because of the non-linear nature of $J_{\mu \nu },\tilde{Q},\sqrt{%
-p^{2}}$ as functions of $\left( J,p,Q\right) $ the resulting
exponential must be a $\diamond $-star exponential. This is
understood as follows. A priori we have started with the classical
$( J,p,Q) $ as the Weyl images of the quantum operators $(
\hat{J},\hat{p},\hat{Q}) .$ Any function of these operators has an
image that is computed by replacing each
$(\hat{J},\hat{p},\hat{Q}) $ by its classical image $(J,p,Q)$
but multiplied with each other by using the $\diamond $%
-product. With that definition, the $J^{\mu \nu }$ and $\tilde{Q}_{\dot{%
\alpha}}$ that appear in the exponent in Eq.(\ref{diamond}) are
constructed by inserting the $\diamond $ product, such as $J^{\mu
\nu }=\left( -p^{2}\right) ^{-1/4}\diamond J^{[\mu }\diamond
p^{\nu ]}\diamond \left( -p^{2}\right) ^{-1/4}+\cdots .$
Furthermore, when these are multiplied to build the exponential
series, one should always use the star product. Then
Eq.(\ref{diamond}) is understood as the image of its corresponding
operator equation.

However, there are some simplifications
that permit us to substitute the classical forms of $J_{\mu \nu },\tilde{Q},%
\sqrt{-p^{2}}$ as mentioned above. First, by noting that $\tilde{Q}_{\dot{%
\alpha}}$ and $S_{\mu \nu }$ are constructed only from
(anti)commuting operators we realize that they cannot have any
corrections
from the deformation parameter\footnote{%
In this paper for convenience we will refer to the deformation parameter as $%
\hbar .$ However if we apply the formalism to describe string
joining/splitting in string field theory, the deformation
parameter is unrelated to the $\hbar $ in quantum mechanics.}
$\hbar $ since one can freely change the orders of $(J,p,Q)$ as
quantum operators in these expressions (thanks to the symmetry
structure of the gamma matrices in $S_{\mu\nu}$ the nontrivial
anticommutator between two $Q$'s does not contribute). Second,
since the expression $\left( -p^{2}\right) ^{-1/4}\diamond J^{[\mu
}\diamond p^{\nu ]}\diamond \left( -p^{2}\right) ^{-1/4}$ is
hermitian, its Weyl image must be real. Therefore it cannot have a
contribution at first order in $\hbar $ (odd orders are
imaginary). It cannot have $\hbar $ corrections to higher orders
either because the star product at higher orders involves higher
order derivatives that vanish on a function that is linear in $J$
(analogous to the differential operator version of Moyal star).
Hence, the images for the nonlinear expressions for $J_{\mu \nu
},\tilde{Q},\sqrt{-p^{2}}$ are simply their classical expressions.

A systematic expansion of the $\diamond $-product in powers of
$\hbar $ can be given as follows. The parameter $\hbar $ comes
from two sources: first the star exponentiation, and second the
coefficients $\left( k_{12}^{\mu },\cdots \right) $ which we can
compute group theoretically to all orders of $\hbar $.  Up to
second order in $\hbar $ only the expansion of the coefficients
$\left( k_{12}^{\mu },\cdots \right) $ contribute. Starting with
the third order the star exponential also contributes. To compute
the contribution from the star exponential, one can use the known
form of the star product at one lower order. In this way one can
obtain systematically a completely explicit form of the $\diamond
$-product to all orders of $\hbar . $ By applying this method we
have verified \textit{\`{a} posteriori} that indeed the
expressions for $J_{\mu \nu },\tilde{Q},\sqrt{-p^{2}}$ do not
receive any contributions from $\hbar .$ Similarly, the statements
in footnote 6 can be verified \textit{\`{a} posteriori}.

The result for the $\diamond $-product may also be written in the
differential form of Eq.(\ref{stardiff}) (with $\partial /{\partial J}$
derivatives, not $\partial /{\partial x}$ derivatives). This last form is
appropriate for computing the star product for polynomials (such as $x^{\mu
}\diamond x^{\nu }$) or any other functions of $A(x,p,Q)$ after writing them
in terms of $(J,p,Q)$. For example, from the group theoretical result in the
$J$-basis, we can also determine the superstar product in the $x$-basis $%
e^{ik_{1}\cdot x+iq_{1}\cdot p+i\psi _{1}\cdot Q}$ by writing it in the $J$
basis as $\left( e^{ik_{1}\cdot J(-p^{2})^{-1/2}+iq_{1}\cdot p+i\psi
_{1}\cdot Q}\right) _{\diamond }$, with a $\diamond $-star exponential.

In the purely bosonic case, the classical basis $e^{ik_{1}\cdot
x+iq_{1}\cdot p}$ is simpler than the basis $\left( \exp \left( ik_{1}\cdot x%
\sqrt{-p^{2}}+iq_{1}\cdot p\right) \right) _{\star }=e^{ik_{1}\cdot
J+iq_{1}\cdot p},$ where $\star $ reduces to the standard Moyal product in
the absence of fermions. Similarly, the $\diamond $-products for the Fourier
basis $e^{ik_{1}\cdot x+iq_{1}\cdot p},$ when expressed in the $\left(
J,p\right) $ space, should give the same result as the Moyal product (using
the Moyal $\star $)%
\begin{equation}
e^{ik_{1}\cdot x+iq_{1}\cdot p}\star e^{ik_{2}\cdot x+iq_{2}\cdot
p}=e^{i\left( k_{1}+k_{2}\right) \cdot x+i\left( q_{1}+q_{2}\right) \cdot
p}~e^{-\frac{i}{2}\left( k_{1}\cdot q_{2}-k_{2}\cdot q_{1}\right) }.
\end{equation}%
Therefore, the following computation is a test of our formalism
\begin{eqnarray}
&&\left( e^{ik_{1}\cdot J(-p^{2})^{-{\frac{1}{2}}}+iq_{1}\cdot p}\right)
_{\diamond }\diamond \left( e^{ik_{2}\cdot J(-p^{2})^{-{\frac{1}{2}}%
}+iq_{2}\cdot p}\right) _{\diamond }  \notag \\
&=&\left( e^{i\left( k_{1}+k_{2}\right) \cdot J(-p^{2})^{-{\frac{1}{2}}%
}+i\left( q_{1}+q_{2}\right) \cdot p}\right) _{\diamond }~e^{-\frac{i}{2}%
\left( k_{1}\cdot q_{2}-k_{2}\cdot q_{1}\right) },
\end{eqnarray}%
where only $\diamond $-exponentials must appear. Indeed this is correct. The
technical details of this computation will be given in another paper \cite%
{BD}.

Thus, as expected, the Moyal product $\star $ in the $\left( x,p\right) $
basis and our group theoretical $\diamond $-product in the $\left(
J,p\right) $ basis are equivalent when the fermions are absent. In the
supersymmetric case it remains to be seen whether one basis is superior to
the other in practical computations (explicit computations are in progress).

By following this program we can compute the superstar product for general
fields%
\begin{equation}
A\left( J,p,Q\right) \star B\left( J,p,Q\right) \;or\;A^{\prime }\left(
x,p,Q\right) \star B^{\prime }\left( x,p,Q\right) .
\end{equation}%
The result appears complicated but it has a completely tractable group
theoretical structure. We hope to give explicit calculations using these
formulas in the near future. Our results as well as methods are likely to be
useful in various applications. In particular, we hope that it can be used
in the formulation of superstring field theory, which was the motivating
factor of our investigation.

In this paper we achieved our main goal of formulating the superstar, but
along the way we also obtained two other new results. First we gave the
quantization of the off shell superparticle in $d$-dimensions and showed
that its quantum mechanics gives a nonlinear realization of $(d+1)$%
-dimensional Poincar\'{e} superalgebra. This higher structure was essential
for constructing the superstar. Second, we introduced efficient group
theoretical methods for constructing and computing star products. We showed
that for the maximum linear Poisson structure, as well as for non-linear
Poisson structures that can be embedded as part of non-linearly realized Lie
(super)algebras, we can obtain the exact full (super)star product by using
group theory representations. This concept is useful for performing explicit
computations involving the superstar in its applications.

\section*{Acknowledgements}

I.B. would like to thank M. Lledo for discussions. I.B. is in part supported
by a DOE grant DE-FG03-84ER40168. He is grateful to the Physics Department
at UC Berkeley, the Feza G\"{u}rsey Institute, and the CERN TH-division, for
hospitality while this work was performed. C.D. is supported in part by the
Turkish Academy of Sciences in the framework of the Young Scientist Award
Program (CD/T\"{U}BA--GEB\.{I}P/2002--1--7). A.P. and B.Z. are supported in
part by the DOE under contract DE-AC03-76SF00098 and in part by the NSF
under grant 22386-13067.

\end{document}